\documentclass[runningheads]{llncs}
\usepackage[T1]{fontenc}
\usepackage{graphicx}
\usepackage{xcolor}
\usepackage{hyperref}

\usepackage{url}

\usepackage{enumitem}
\usepackage{tabularx}
\usepackage{threeparttable} 

\usepackage{color,soul} 
\sethlcolor{yellow}
\usepackage{subcaption}
\usepackage{multirow}
\usepackage{amssymb}

\usepackage{booktabs}

\usepackage{xspace}

\newcommand{\name}{SecMate} 
\newcommand{\namebas}{VCA$^{Baseline}$\xspace}
\newcommand{\nameadap}{\name$^{Adap}$\xspace}
\newcommand{\nameCC}{\name$^{CC}$\xspace}
\newcommand{\nameboth}{\name$^{Both}$\xspace}
\newcommand{\namenone}{\name$^{None}$\xspace}

\begin{document}
\title{\name: Multi-Agent Adaptive Cybersecurity Troubleshooting with Tri-Context Personalization}
\titlerunning{\name: Multi-Agent Adaptive Cybersecurity Troubleshooting}
%
\author{Yair Meidan\inst{1}\orcidID{0000-0003-4865-2334} \and
Omri Haller\inst{1}\orcidID{0009-0002-1140-7167} \and
Yulia Moshan\inst{1}\orcidID{0009-0003-0775-0939} \and
Shahaf David\inst{1}\orcidID{0009-0003-9935-5230}  \and
Dudu Mimran\inst{1}\orcidID{0009-0004-9610-6156}\and Yuval Elovici\inst{1}\orcidID{0000-0002-9641-128X} \and \\ Asaf Shabtai\inst{1}\orcidID{0000-0003-0630-4059}}
\authorrunning{Y. Meidan et al.}
%
\institute{Ben-Gurion University of the Negev
}
\maketitle              
\begin{abstract}
Recent advances in large language models and agentic frameworks have enabled virtual customer assistants (VCAs) for complex support.
We present \name{}, a multi-agent VCA for cybersecurity troubleshooting that integrates device, user, and service specificity from conversational and device-level signals.
Device specificity is provided by a lightweight local diagnostic utility, while user specificity relies on implicit proficiency inference and profile-aware troubleshooting.
Service specificity is achieved through a proactive, context-aware recommender.
We evaluate \name{} in a controlled study with 144 participants and 711 conversations.
Device-level evidence increased correct resolutions from about 50\% to over 90\% relative to an LLM-only baseline, while step-by-step guidance improved pleasantness and reduced user burden.
The recommender achieved high relevance (MRR@1 $\approx$ 0.75), and participants showed strong willingness to substitute human IT support at costs well below human benchmarks.
We release the full code base and a richly annotated dataset to support reproducible research on adaptive VCAs.

\keywords{Agentic AI \and Chatbots \and Cybersecurity Troubleshooting.}

\end{abstract}
\section{Introduction}\label{sec:Introduction}

Virtual customer assistants (VCAs) increasingly replace human support representatives and are expected to continue doing so at scale~\cite{fortune2025salesforce}.
In cybersecurity, users spanning diverse ages, educational backgrounds, and technical proficiency levels frequently require assistance with tasks such as diagnosing abnormal system behavior or assessing the risk of potentially malicious files.
In organizational contexts, particularly within small and medium sized businesses (SMBs), internal cybersecurity support is often unavailable, leading employees to rely on external managed security service providers (MSSPs)~\cite{morris2023cybersecurity}.

\begin{figure}[t]
    \centering
    \includegraphics[width=0.99\columnwidth,clip,trim=0.7cm 0.25cm 0.7cm 0.2cm]{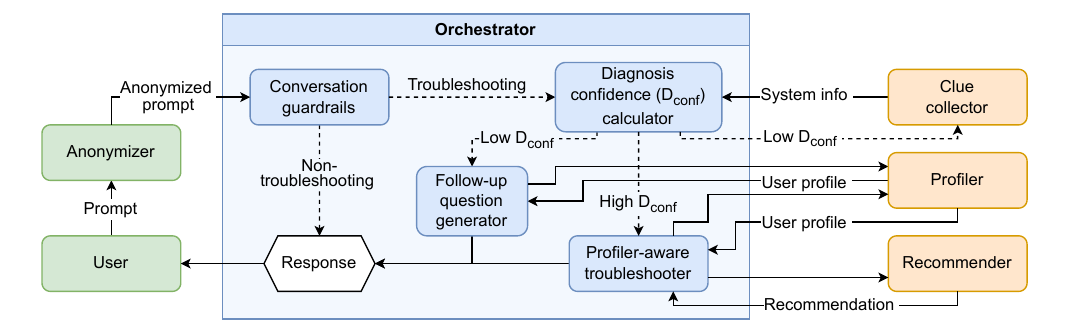}
    \caption{Per-iteration troubleshooting workflow of \name{}.  
Solid arrows denote unconditional steps, while dashed arrows indicate logic-dependent paths.
}
    \label{fig:orchestrator_overview}
\end{figure}

To address the scalability limitations of MSSPs and to improve the troubleshooting experience of SMB users, we introduce \name, a device-, user-, and service-specific multi-agent VCA for cybersecurity troubleshooting, that is grounded in pattern recognition from conversational and device level signals.
As illustrated in Fig.~\ref{fig:orchestrator_overview}, \name{} coordinates specialized agents to guide users through troubleshooting in a step by step manner while balancing efficiency, diagnostic accuracy, and user satisfaction.
With explicit user consent, a \emph{Clue Collector} (CC) agent supplies local system evidence to ground the conversation and reduce diagnostic ambiguity.
A \emph{Profiler} agent implicitly and continuously infers the user’s technical proficiency, enabling a \emph{Profile-Aware Troubleshooter} agent to translate complaints and contextual clues into tailored diagnostic paths and corrective actions.
When diagnostic confidence remains insufficient, a \emph{Follow-Up Question Generator} agent produces targeted, profile-aware queries to elicit additional information from the user.
Complementing the core troubleshooting workflow, a proactive \emph{Recommender} agent suggests relevant products that support resolution or prevention, contributing to \name{}'s long term sustainability.

We implemented \name{} and conducted a large-scale user study (approved by our institution's Human Subjects Research Committee).
Results show that device-level evidence collection increased task success from about 50\% to over 90\%, while step-by-step, profile-aware guidance lowered user burden.
Accurate implicit profiling improved communication quality, solution ordering, and technical appropriateness, whereas misprofiling sharply degraded perceived quality.
The Recommender achieved high relevance (MRR@1 $\approx$ 0.75) with positive perceptions, revealing a trade-off between recommendation timing and conversational smoothness.
Users preferred incremental solution delivery and expressed willingness to substitute human IT support with \name{}.
Conversation costs stayed well below human support benchmarks, indicating economic viability.

\textbf{Contributions:}
(1) We present a novel multi-agentic orchestration mechanism that adapts cybersecurity support based on tri-context data, integrating device-grounded evidence, implicit user profiling, and context aware recommendation, moving beyond prior work (Sec.~\ref{sec:background_and_related_work}) that addresses these components in isolation.
(2) We operationalize and extend prior methods for profiling~\cite{david2025profillm} and recommendation~\cite{haller2025impress} by embedding them in an interactive troubleshooting pipeline and evaluating their system level impact, rather than standalone accuracy, through a large scale online user study.
(3) We introduce the CC, a device grounding agent that transforms a general purpose large language model (LLM) into a device specific troubleshooter by incorporating real time local evidence into the conversational loop.
(4) We release\footnote{Code and data will be released upon paper acceptance.} code and a richly annotated dataset of 711 real user conversations from 144 participants across five cybersecurity scenarios, enabling reproducible research on adaptive, personalized conversational systems and their user experience tradeoffs.


\section{Proposed Method}\label{sec:proposed_method}

We formulate cybersecurity troubleshooting as a sequential inference problem, whose objective is to iteratively infer the most plausible diagnostic and remediation path from heterogeneous evidence.  
The considered evidence sources include natural language user utterances, structured device diagnostics, and an evolving latent user proficiency profile.  
This formulation enables unified reasoning over heterogeneous inputs and allows \name{} to adapt the content, ordering and phrasing of troubleshooting actions, thereby improving support effectiveness and user satisfaction.  
Below, we describe \name{}’s key agentic components and the overall workflow coordinated by the Orchestrator.  
Additional implementation details, including the exact prompts used, are available in our code repository\footnote{To be released upon paper acceptance.}.

A \name{} iteration (from user prompt to VCA response) begins with prompt anonymization, preserving privacy by de-identifying personally identifiable information (PII)~\cite{microsoft_presidio}.  
Once guardrails confirm troubleshooting intent, the anonymized prompt is processed by task-specific agents coordinated by the Orchestrator.

\textbf{Clue Collector (CC):}  
The CC relies on a lightweight local utility that gathers system information such as OS version, active processes, resource use, installed software and hardware, connected peripherals, and network configuration.  
This evidence supports efficient root-cause analysis and reduces user effort, for example by avoiding complex instructions.  
Use of the CC is optional and requires user consent; when activated, it collects endpoint data from the start of the conversation to ensure immediate availability.  
The Orchestrator determines when and whether to retrieve this evidence based on its informative value.

\textbf{Follow-Up Question Generator:}  
User prompts or CC evidence may be insufficient to isolate the root cause or confirm a solution.  
Common gaps include vague descriptions, missing details about prior actions, or app-specific error messages unavailable to the CC.  
The LLM-based Follow-Up Question Generator identifies informative missing details and asks targeted questions, e.g., “From where did you download this file?”  
As shown in Fig.~\ref{fig:orchestrator_overview}, it triggers the Profiler to update the user’s technical profile, and phrases follow-up questions accordingly.

\textbf{User Profiler:}  
To profile VCA users’ IT/cybersecurity troubleshooting proficiency, we adopt the ProfiLLM taxonomy and method~\cite{david2025profillm}.  
The taxonomy comprises 23 subdomains covering common IT/cybersecurity issues for which PC users seek troubleshooting assistance~\cite{MicrosoftDDR2023}, grouped into five domains.  
User proficiency is represented as a 23-dimensional vector with scores ranging from one (low) to five (high).  
In each iteration, the Profiler identifies relevant subdomains, assigns temporary scores, and updates profile values via weighted averaging.
While ProfiLLM was originally optimized offline for profiling accuracy using conversations with 80 users, in \name{} we evaluate its impact on profile-aware troubleshooting and system-level performance when embedded in an interactive VCA pipeline, using interactions with nearly twice as many online participants.

\textbf{Profile-Aware Troubleshooter:}
This LLM-based agent leverages the inferred user profile to (1) prioritize diagnostic and corrective actions, and (2) adapt VCA responses, including follow-up questions, for example by adjusting jargon and detail level.
Given user prompts, CC evidence, and the inferred profile, \name’s Troubleshooter selects the most \emph{suitable} (rather than merely the most probable) diagnostic path.
For example, for users with limited IT/cybersecurity proficiency reporting PC slowness, the Troubleshooter suggests basic actions such as restarting the PC, instead of requesting advanced inspections (e.g., terminating suspicious processes via the Task Manager).
If the issue persists, the Troubleshooter advances to more complex diagnostic and remediation steps.

\textbf{Proactive Recommender:}  
\name{} proactively suggests relevant products during troubleshooting using ImpReSS’s implicit recommendation mechanism~\cite{haller2025impress}.  
This supports effective troubleshooting, cybersecurity problem solving and prevention, and MSSP revenue generation, while minimally disrupting the conversation flow.  
The Recommender continuously analyzes the dialogue, retrieves candidate products, and ranks them by relevance to the inferred diagnosis.  
We extend ImpReSS in four ways.  
First, we evaluate the Recommender by its impact on an end-to-end, recommendation-augmented VCA rather than standalone accuracy.  
Second, we empirically compare presentation strategies, including in-chat and pop-up delivery.  
Third, we analyze large-scale human-VCA conversations instead of synthetic or small datasets.  
Fourth, we add a reasoning layer in which an LLM generates a brief, context-dependent justification for each recommendation.  
When a recommendation is triggered, the Profile-Aware Troubleshooter integrates the rationale-backed suggestion inline or via a popup.

\textbf{Confidence-Guided Orchestrator:}
Preliminary experiments with commercial LLMs (e.g., ChatGPT) revealed recurring limitations for self-service cybersecurity support, including overly verbose responses with multiple parallel diagnoses, limited follow-up questions, overconfident resolutions, and interaction styles that diverge from human support agents.
To address these issues, we introduce the Orchestrator, whose objectives are to (1) optimize troubleshooting dialogues for user satisfaction and (2) dynamically coordinate \name{}’s components to provide efficient, device-, user-, and service-specific support.

At each iteration, the Orchestrator analyzes the conversation history and current complaint to select the next action: generating a solution, asking a follow-up question, or retrieving device evidence from the CC. When producing solutions or questions, it first consults the Profiler to adapt content to the user’s expertise. This decision is guided by a \emph{diagnosis confidence} score ($D_{conf}$), computed by an LLM based on evidence strength, the diversity of plausible diagnoses, and prior conversational outcomes.
If $D_{conf}$ is low, the Orchestrator requests additional evidence from the CC or elicits targeted user input. If $D_{conf}$ is sufficiently high, it invokes the Profile-Aware Troubleshooter, segments the solution into concise steps, and presents them sequentially. After each step, users may proceed or request clarification, enabling controlled, human-like troubleshooting progression.

\section{Related work}\label{sec:background_and_related_work}

\textbf{Limitations of state-of-the-art LLMs.}
Preliminary experiments show that leading LLMs (e.g., ChatGPT, Gemini, Claude) can map user complaints to plausible cybersecurity diagnoses and remediation steps.
However, they typically generate verbose responses with unprioritized alternatives, overwhelming users and rarely posing follow-up questions to disambiguate competing diagnoses.

\textbf{Virtual customer assistants.}
VCAs have evolved from template-driven systems based on FAQs and decision trees~\cite{ngai2021intelligent,10841729} to ML- and LLM-based approaches supporting intent recognition, contextual modeling, and dialogue~\cite{10574658,boonyingdevelopment,ma2025multimodal}.
Agentic and agentic tool-orchestrating (AGTO) systems extend VCAs by coordinating reasoning, retrieval, and tool execution via specialized agents~\cite{abbasian2025conversational,10393016,kaheh2023cyber,kiangala2024experimental}.
Most prior work addresses narrow tasks (e.g., log analysis, clinical diagnostics, industrial troubleshooting) and is evaluated on small or synthetic datasets.
Commercial systems such as Microsoft Copilot for Service~\cite{Microsoft2025} and Salesforce Agentforce~\cite{Salesforce2025} 
emphasize enterprise integration and staff augmentation.
In contrast, \name{} targets self-service cybersecurity support and jointly models device evidence, user proficiency, and product recommendation within a unified agentic architecture, evaluated in a large-scale controlled study.

\textbf{Data sources for personalized VCA support.}
Existing systems leverage heterogeneous data sources, including organizational FAQs~\cite{ngai2021intelligent}, industrial telemetry~\cite{kiangala2024experimental}, and multimodal clinical data~\cite{abbasian2025conversational,ma2025multimodal}.
\name{} integrates two complementary evidence streams within a single inference pipeline: user-generated conversational input and real-time device-level signals collected by the CC agent.

\textbf{Device specificity.}
AGTO systems increasingly incorporate device-level sensing to ground inference in evidence.
Prior work demonstrates benefits in healthcare~\cite{abbasian2025conversational,ma2025multimodal}, industrial maintenance~\cite{kiangala2024experimental}, IT support~\cite{boonyingdevelopment}, and cybersecurity operations~\cite{10574658,10393016,kaheh2023cyber}.
\name{} extends this paradigm to end users by using the CC to capture transient system states (e.g., running processes, open ports, installed software) and inject them into the troubleshooting loop in real time.
This reduces reliance on self-reporting and grounds diagnostic inference in verifiable context, similar to systems that reason over fragmented clues~\cite{wang2025cluecart}.

\textbf{User specificity.}
Prior personalization approaches often require users to adapt their behavior or prompts~\cite{Yeh2022How}, or focus on stylistic adaptation such as tone, empathy, or anthropomorphism~\cite{uzan2025personalization,xu2023anthropomorphic}.
While effective in healthcare and education~\cite{kim2023healthcare,zhou2022adaptive}, these methods primarily adapt communication style.
In contrast, \name{} performs implicit user profiling to infer IT/cybersecurity proficiency and adapts the content and complexity of diagnostic actions accordingly.

\textbf{Service specificity.}
Conversational recommender systems typically assume explicit seeker-recommender interactions~\cite{Bernard_2023}, emphasizing retrieval~\cite{yang2024unleashing}, knowledge graph reasoning~\cite{ma-etal-2021-cr}, or persuasion~\cite{Bernard_2023}.
\name{} instead embeds an \emph{implicit} recommender into cybersecurity troubleshooting, where recommendations are inferred from diagnostic context rather than purchasing intent.
We evaluate this component in a controlled study, measuring retrieval accuracy with perceived fit, clarity, and integration, and comparing multiple recommendation presentation strategies, a scope largely absent from prior cybersecurity support work.

\section{Evaluation method}\label{sec:Evaluation_method}

\textbf{Secure and privacy-preserving experimental setup.}
We implemented \name{} as a distributed Python system, using LangChain as the LLM wrapper, LangGraph for agent orchestration, and LangSmith for observability.
To support modularity and scalability throughout the user study, components were deployed as independent microservices.
The conversational core relied on OpenAI’s GPT-4o, paired with a React frontend and hosted on AWS using Route~53 for DNS and load balancing, while security and privacy constraints guided the architecture: client-server communication was encrypted via HTTPS with JSON Web Tokens (JWT), the CC communicated over WebSocket Secure (WSS), authentication was handled by Amazon Cognito, all services ran inside a private Virtual Private Cloud (VPC) with restricted subnets and security groups, and PII was processed using the Microsoft Presidio library~\cite{microsoft_presidio} to mitigate leakage risk.

\textbf{User study protocol.}  
Following institutional ethics approval, we conducted a controlled VCA user study and recruited participants progressively, incorporating minor refinements in response to feedback from early sessions.  
In each session, participants first provided informed consent, optionally reported demographic information, and completed two online questionnaires used as ground-truth (GT) labels for IT/cybersecurity proficiency\footnote{URL to be included upon acceptance} and personality traits\footnote{https://openpsychometrics.org/tests/IPIP-BFFM/}.  
Each session began with a five-minute warm-up conversation, during which participants interacted freely with \name{} by discussing their cybersecurity background or related experiences.  
Participants then used \name{} to troubleshoot five predefined cybersecurity scenarios (Table~\ref{tab:scenarios}), simulated on the experiment laptops.

\begin{table*}[ht]
\centering
\caption{The cybersecurity-related troubleshooting scenarios in our user study.}
\label{tab:scenarios}
\tiny 
\begin{tabularx}{\textwidth}{p{0.8cm} p{1.4cm} p{1.7cm} p{1.6cm} p{2.3cm} p{2.4cm}}
\toprule
\multicolumn{1}{c}{\textbf{Title}} & \multicolumn{1}{c}{\textbf{User Complaint}} & \multicolumn{1}{c}{\parbox{2.1cm}{\centering \textbf{Simulated \\ Scenario}}} & \multicolumn{1}{c}{\parbox{2.3cm}{\centering \textbf{Experiment \\ Implementation}}} & \multicolumn{1}{c}{\textbf{Expected Outcome}} & \multicolumn{1}{c}{\textbf{Relevant SPCs}} \\ 
\midrule
PC performance & My PC is slow. & Cryptominer~\cite{varlioglu2022dangerous} consumes system resources. & Memory-intensive app (Prime95~\cite{prime95}) runs silently. & Identify Prime95 as root cause; possibly end task & Device utility SW, PC cleaner SW, endpoint detection and response (EDR) \\ 
Online gaming & Child may have clicked something in a game. & Adware~\cite{szatmary2024cybersecurity} lures child to download malicious .exe. & fortnite\_cheats.exe in Downloads, not executed. & Identify fortnite\_cheats.exe as potentially malicious; possibly delete & Malware scanners, antivirus, EDR \\ 
Airport Wi-Fi & Is it safe to access my bank on airport Wi-Fi? & Risks of public airport Wi-Fi~\cite{cheng2013characterizing}. & PC connected to open Wi-Fi network. & Identify open Wi-Fi as insecure; possibly use hotspot, VPN, or secured network & Virtual private network (VPN), mobile hotspot, secure web gateway, internet security suite \\ 
Safe PC & Is my PC safe to use? & Ensuring the PC is free of security issues. & Firewall is disabled. & Identify MS Defender firewall disabled; possibly enable it & Standalone firewall, internet security suite, antivirus, EDR \\ 
Moving cursor & My mouse moved once by itself; was I hacked? & Hidden remote access Trojan~\cite{valeros2020growth} installed. & TeamViewerQS~\cite{teamviewerqs} in Installed Apps; installer in Downloads. & Identify TeamViewerQS installed; possibly uninstall and delete & Next-gen antivirus, malware scanners, managed detection and response, EDR \\ 
\bottomrule

\end{tabularx}
\begin{tablenotes}
\tiny
\item \textbf{SPC} = Solution product category (Sec.~\ref{sec:Evaluation_method}).
\end{tablenotes}
\end{table*}

\begin{table*}[ht]
\centering
\caption{Metrics used in our experiments.}
\label{tab:metrics}
\tiny
\begin{tabularx}{\textwidth}{p{0.2cm} p{1.5cm} p{1.5cm} p{6.2cm} p{1.6cm}}
\toprule
\multicolumn{1}{c}{\parbox{1.1cm}{\centering \textbf{Related \\ Entities}}} & \multicolumn{1}{c}{\textbf{Metric}} & \multicolumn{1}{c}{\textbf{Type, Range}} & \multicolumn{1}{c}{\textbf{Description / Calculation}} & \multicolumn{1}{c}{\textbf{Data Source}} \\ 
\midrule
\multirow[c]{10}{*}{\shortstack[l]{System, \\ CC}} & Overall pleasantness & Numeric, discrete {[}1, 5{]} & Overall, I had a pleasant experience using the chatbot & Scenario-level feedback form \\ 
 & Ease of use & Numeric, discrete {[}1, 5{]} & The chatbot was easy to use and interact with & Scenario-level feedback form \\  
 & Efficiency & Numeric, discrete {[}1, $\infty${]} & Number of iterations to reach correct solution: Index of first correct diagnosis or solution, -1 if \emph{the VCA} fails & Conversation log annotation
 \\ 
 & {\shortstack[l]{Effective- \\ ness}} & Boolean {[}0, 1{]} & Did the conversation reach the expected outcome (True=1, False=0), elaborated on in Table~\ref{tab:scenarios} & Conversation log annotation \\ 
 & Perceived effectiveness & Numeric, discrete {[}1, 5{]} & The chatbot effectively addressed my concerns and helped me resolve them & Scenario-level feedback form \\ 
 & {\shortstack[l]{Overwhelm- \\ ingness}} & Numeric, discrete {[}0, $\infty${]} & Number of distinct paths to diagnosis/solution & Conversation log annotation \\ 
 & Perceived overwhelmingness & Numeric, discrete {[}1, 5{]}$\ast$ & If the scenario was "PC Performance" or "Moving cursor" - answer this question. Otherwise mark 0. The number of possible diagnoses provided by the chatbot was (1) too few, (3) just right for me, (5) too many & Scenario-level feedback form \\ 
 & Diagnosis presentation preference & Numeric, discrete {[}1, 5{]} & In troubleshooting scenarios, regarding the \emph{diagnosis} phase, I prefer (1) being guided through potential diagnoses one step at a time to identify the cause, (5) seeing a complete overview of all potential diagnoses upfront & System-level feedback form \\ 
 & Solution presentation preference & Numeric, discrete {[}1, 5{]} & In troubleshooting scenarios, regarding the \emph{solution} phase, I prefer (1) being guided through solution steps one at a time, (5) seeing a complete overview of all steps upfront
 & System-level feedback form \\  
 & Substitution willingness & Numeric, discrete {[}1, 5{]} & In the event of a cybersecurity issue, how likely would you be to use this troubleshooting chatbot instead of a human IT representative? (1) strongly disagree, (5) strongly agree & System-level feedback form \\ \midrule
\multirow{2}{*}{\parbox{1.2cm}{Profiler, Troubleshooter}} & MAE & Numeric, continuous {[}0, 5{]} & Difference between inferred and ground-truth profile vectors & Questionnaires, inferred profile \\ 
 & Wording adequacy & Numeric, discrete {[}1, 5{]} & The chatbot's wording and terminology were (1) too simple, (3) just right for me, (5) too complex & Scenario-level feedback form \\ \midrule
\multirow{4}{*}{\parbox{1.2cm}{Recom\-mender}} & MRR@$k$ & Numeric, continuous {[}0, 1{]} & Mean reciprocal rank at rank $k$: Proportion of cases (scenarios) where top-$k$-ranked items are correct & Correct SPCs (Table~\ref{tab:scenarios}) \\  
 & Perceived rec. correctness & Numeric, discrete {[}1, 5{]}$\ast$ & The recommended product seems like a good fit to address the issue we've discussed & Scenario-level feedback form \\  
 & Rec. clarity & Numeric, discrete {[}1, 5{]}$\ast$ & I understood why the product was recommended in relation to the problem we were discussing & Scenario-level feedback form \\  
 & Rec. smoothness & Numeric, discrete {[}1, 5{]}$\ast$ & To me, the product recommendation felt natural and didn't disrupt the conversation flow & Scenario-level feedback form \\ 
\bottomrule
\end{tabularx}
\begin{tablenotes}
\footnotesize
\item \textbf{$\ast$} = optional zeros for irrelevant questions (e.g., when no recommendation appeared).\\
\end{tablenotes}
\end{table*}

\textbf{Evaluation metrics.} Participants completed a scenario-level feedback form after each scenario and a final questionnaire at the experiment’s end.
Conversation logs were annotated according to predefined guidelines (App.~\ref{app:log_annotation_guidelines}).
These data were used to derive multiple evaluation metrics (Table~\ref{tab:metrics}) capturing system performance and user acceptance, grounded in established empirical frameworks~\cite{davis1989perceived,iso19989241,nielsen1990heuristic,pu2012evaluating}.
To mitigate central and extreme tendency biases~\cite{fischer2004standardization}, Likert-scale responses ([1,5]) were normalized into within-respondent $z$-scores by mean-centering and variance scaling, enabling aggregation across participants and supporting parametric statistical tests in subsequent analyses (Sec.~\ref{sec:Experimental_results}).

\begin{figure}[t]
    \centering   \includegraphics[width=1\columnwidth,clip]{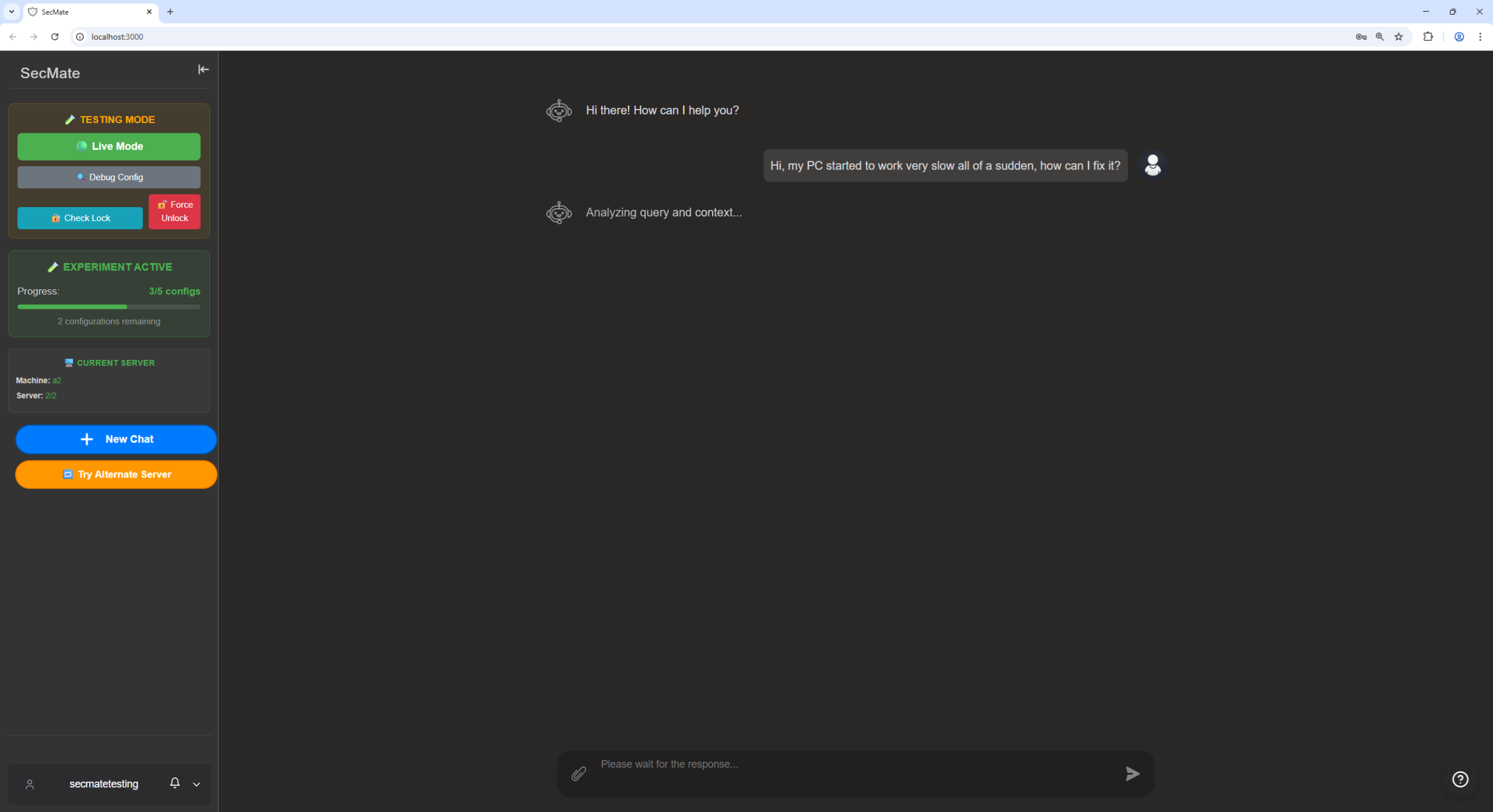}        
\caption{\name{}'s UI, shared across all configurations in our experiments.}
\label{fig:ui}
\end{figure}

\begin{figure}[!h]
  \centering
  \begin{minipage}[b]{0.32\textwidth}
    \vspace*{\fill} 
    \begin{subfigure}{\linewidth}
      \centering
      \includegraphics[width=\linewidth]{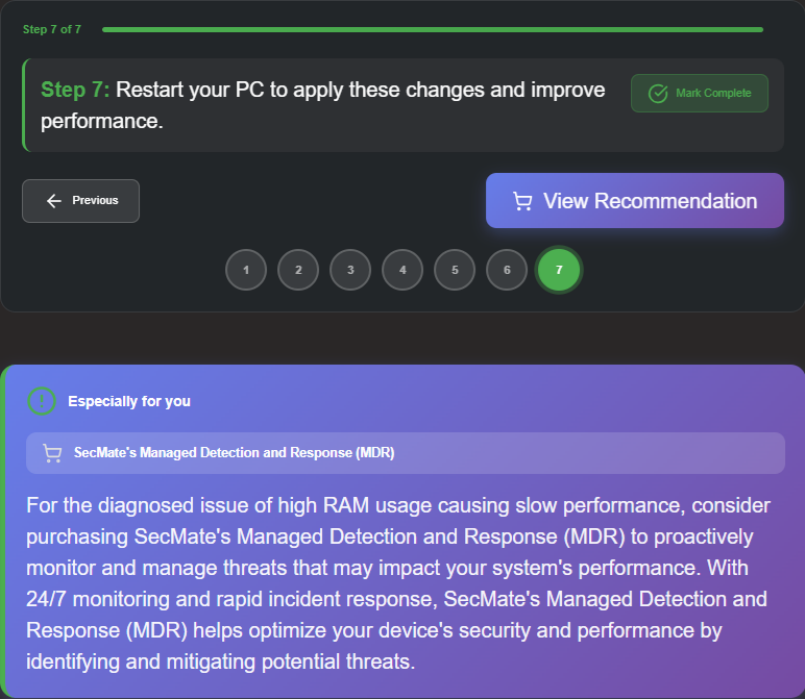}
      \caption{In-chat.}
      \label{fig:in_chat}
    \end{subfigure}
  \end{minipage}\hfill
  \begin{minipage}[b]{0.32\textwidth}
    \vspace*{\fill} 
    \begin{subfigure}{\linewidth}
      \centering
      \includegraphics[width=\linewidth]{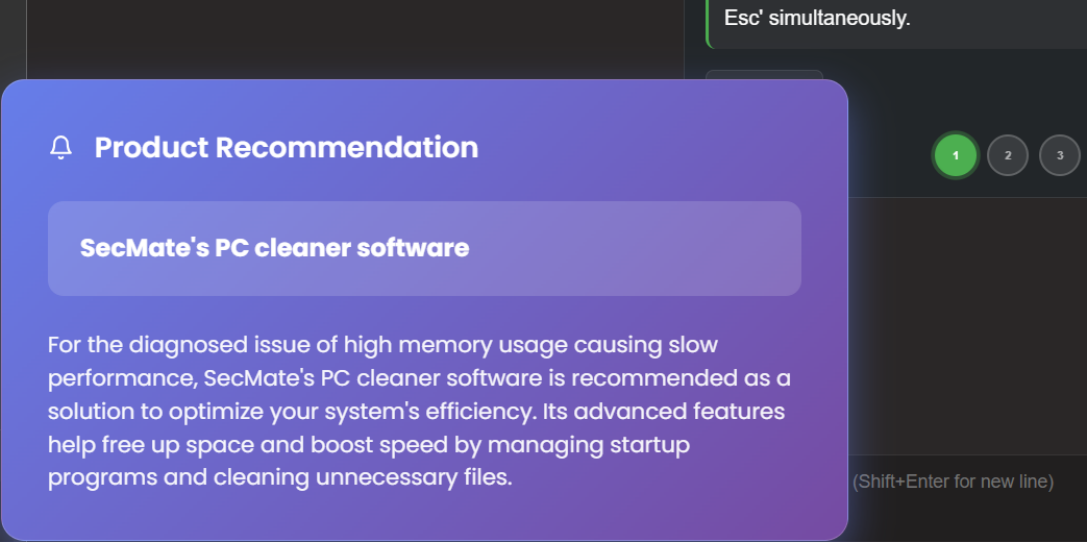}
      \caption{Fixed popup.}
      \label{fig:fixed_popup}
    \end{subfigure}
  \end{minipage}\hfill
  \begin{minipage}[b]{0.32\textwidth}
    \begin{subfigure}{\linewidth}
      \centering
      \includegraphics[width=\linewidth, trim=0.1cm 2.9cm 0.4cm 0cm, clip]{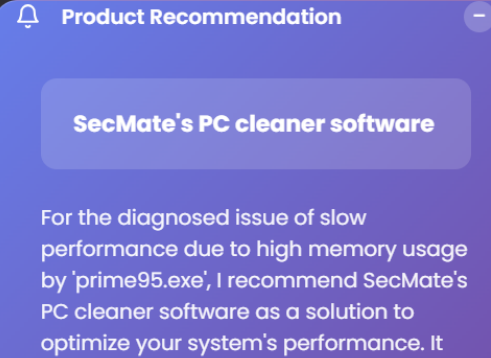}
      \caption{Minimizable popup.}
      \label{fig:minimazable_popup}
    \end{subfigure}
    \begin{subfigure}{\linewidth}
      \centering
      \includegraphics[width=\linewidth]{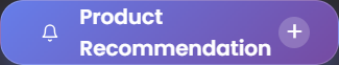}
      \caption{Minimizable popup, collapsed.}
      \label{fig:minimazed_popup}
    \end{subfigure}
  \end{minipage}

  \caption{\name{} recommendation strategies: (a) in-chat, (b) fixed popup, (c) minimizable popup, (d) collapsed minimizable popup.}
  \label{fig:recommendation_presentation_strategies}
\end{figure}

\textbf{Experimental design.}
We evaluated four GPT-4o-powered \name{} configurations covering all combinations of CC enablement and profile-based response adaptation: adaptation only (\nameadap), CC only (\nameCC), both enabled (\nameboth), and neither enabled (\namenone), all using \name{}’s agentic flow and recommendation mechanism.
We also implemented a GPT-4o baseline VCA (\namebas{}) prompted identically to \name{}, which maintains conversational context and provides recommendations but omits agentic orchestration, profile-aware adaptation, and CC-based device evidence.
To prevent bias and preserve internal validity across ablations, all configurations, including \namebas{}, shared the same UI 
(Fig.~\ref{fig:ui}) and recommendation presentation strategy (Fig.~\ref{fig:recommendation_presentation_strategies}).
For each participant, the configuration–scenario mapping and scenario order were randomized, with all participants troubleshooting the same five cybersecurity scenarios without awareness of the configuration.
The study followed a repeated-measures design with randomized scenario-configuration assignment.
We analyzed all metrics using linear mixed-effects models, with configuration as a fixed effect (\namebas{} as reference) and participant and scenario as random effects.
After within-participant $z$-score normalization, effects are reported in standard deviation units rather than the original [1,5] Likert scale.

\section{Experimental Results}\label{sec:Experimental_results}

The study involved 144 engineering students (69\% male, 31\% female), with a mean age of $25.0\pm2.6$ years and $15.3\pm1.3$ years of education.
We denote the dataset of 711 troubleshooting conversations, associated demographics, and profile labels (excluding the introductory five-minute conversations in Sec.~\ref{sec:Evaluation_method}) as $DS_{Complete}$.
Before presenting component-, system-, and overhead-level results, we describe dataset refinements driven by experimental, operational, and technical considerations, as well as participant feedback during the study.

\textbf{Dataset refinements.}
Although the CC collected device evidence in all \name{} configurations, the orchestrator could retrieve it only in \nameCC and \nameboth, and only when deemed informative.
When the CC was never accessed in a conversation, we retrospectively relabeled \nameCC as \namenone and \nameboth as \nameadap to isolate the CC’s effective contribution.
Conversely, when technical failures prevented user profile retrieval, response adaptation could not occur, and we relabeled \nameadap as \namenone and \nameboth as \nameCC.
Applying these rules to $DS_{Complete}$ yielded the $DS_{Relabeled}$ dataset (detailed counts are reported in App.~\ref{app:Dataset_Refinement_Process}).
To assess participant attentiveness, we intentionally injected incorrect profiles or SPC recommendations into a subset of conversations.
These cases were analyzed separately, while all other analyses exclude them to reduce noise and isolate experimental effects, yielding $DS_{Filtered}$.
Finally, adjusting the CC refresh rate to five seconds and making popup recommendations user-minimizable yielded $DS_{Refined}$, which we use in most final analyses as the cleanest \name{} configuration.

\begin{figure}[!b]
    \centering
    \includegraphics[width=0.98\columnwidth,clip,trim=0.1cm 46cm 0.4cm 0cm]{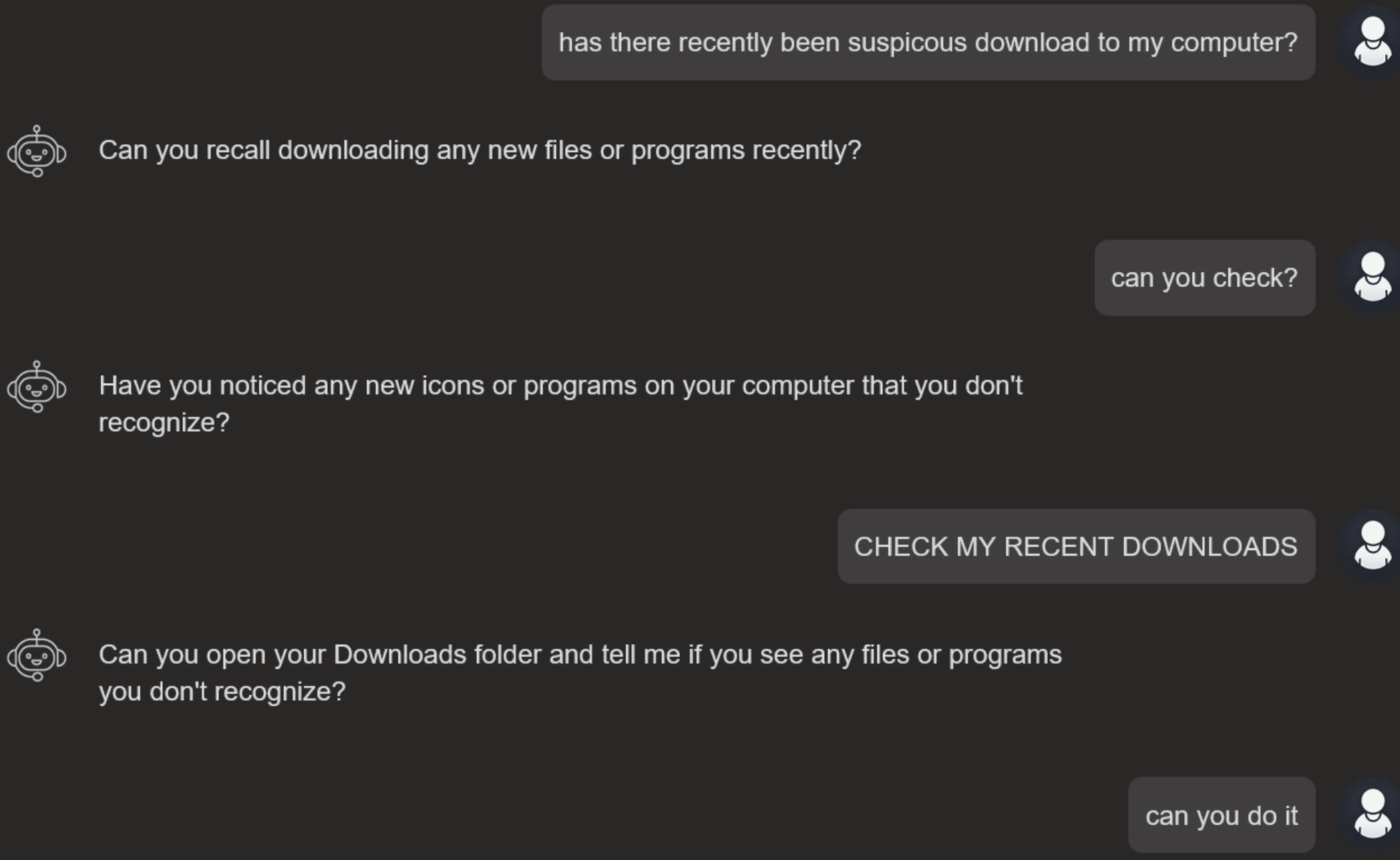}
    \caption{Conversation snippet where the user resists manual inspection and expects \name{} to inspect the PC directly, demonstrating a user's expectation of a device-specific VCA's behavior.}
    \label{fig:user_insists_on_CC}
\end{figure}

\textbf{Clue collection.}
LME analysis on $DS_{Refined}$, using CC-disabled \name{} configurations as baseline, showed high overwhelmingness without the CC ($M$=4.21, $p<0.001$)\footnote{LME effects are reported in within-participant SD units via per-participant $z$-scoring of Likert responses (Sec.~\ref{sec:Evaluation_method}). $M$ denotes the intercept and $\Delta M$ condition effects.}.
Enabling the CC reduced overwhelmingness by limiting diagnostic and solution paths ($\Delta M$=-0.76 SD, 95\% CI [-3.05, 1.53]).
Although this effect was not statistically significant, qualitative inspection revealed multiple cases where users actively requested device-specific support, emphasizing its perceived value in reducing diagnostic effort (Fig.~\ref{fig:user_insists_on_CC}).
\emph{Perceived} overwhelmingness showed a similar pattern, with no significant CC effect ($M$=2.88 SD, $\Delta M$=0.08 SD).
We attribute these null effects to uniform prompting across configurations to provide a single polite solution.
In contrast, CC enablement substantially improved effectiveness: correct solutions were reached in 90.9\% of conversations with the CC, compared to 50\% without it ($\chi^2$=12.46, $p<0.001$).
Compared to single-shot collection, near-real-time CC refresh further increased pleasantness by 0.55 SD ($p$=0.017).
Perceived effectiveness improved non-significantly ($\Delta M$=0.34 SD), while the single-shot baseline was marginally negative ($M$=-0.24 SD), indicating residual shortcomings.

\begin{figure}[ht]
  \centering
  \begin{minipage}{0.46\textwidth}
    \centering
    \includegraphics[width=\linewidth,trim=0.2cm 1.5cm 0.5cm 0.3cm,clip]{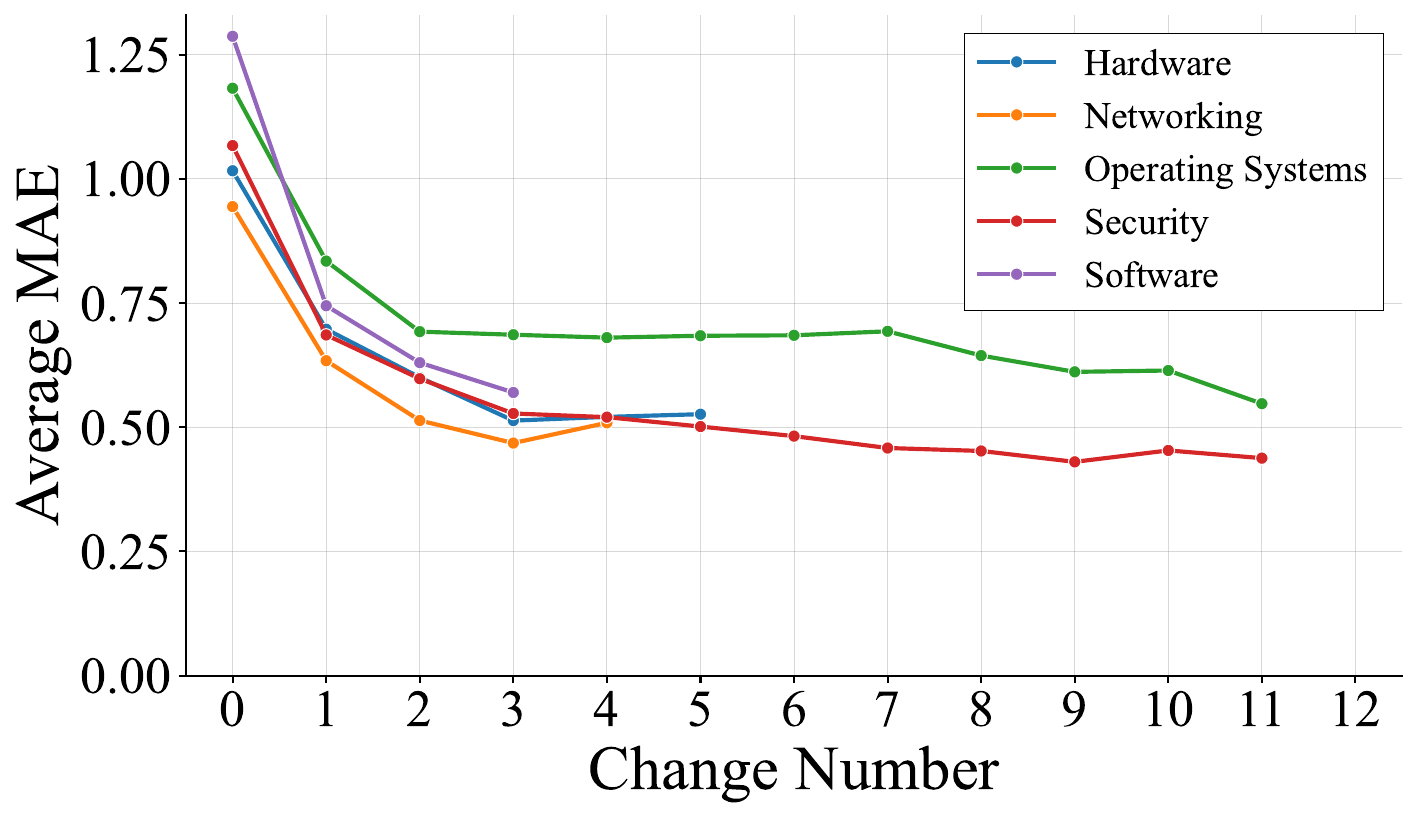}
    \caption{Profile inference accuracy, expressed as average MAE across domains, improving over successive conversational iterations.}    
    \label{fig:domain_avg_MAE}
  \end{minipage}\hfill
  \begin{minipage}{0.51\textwidth}
    \centering    \includegraphics[width=\linewidth,trim=0.2cm 0.7cm 0.8cm 0.4cm,clip]{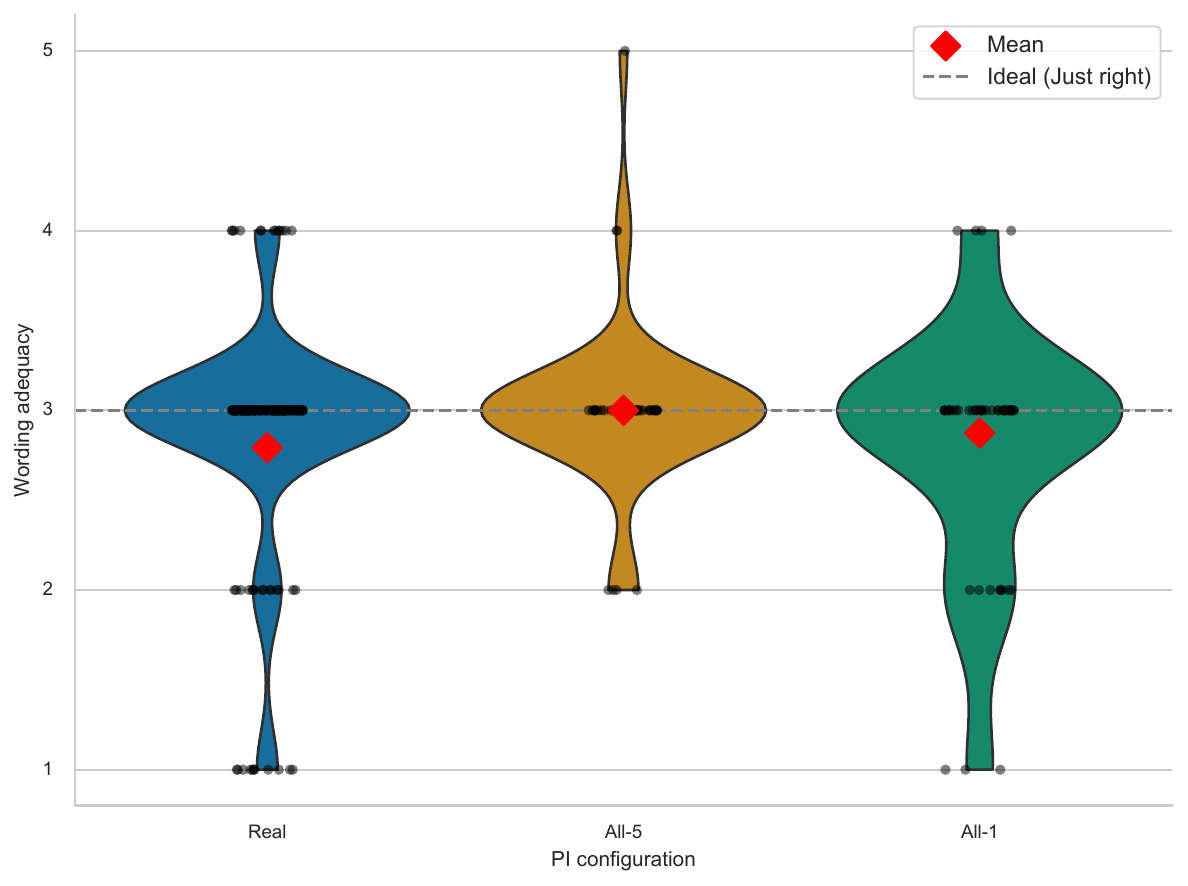}
    \caption{Wording adequacy (original Likert scale) under correct profile-based adaptation versus intentionally incorrect profiles (All-5: expert, All-1: novice).}        \label{fig:wording_clarity_real_1one_five}
  \end{minipage}
\end{figure}

\begin{figure}[ht]
  \centering
  \begin{subfigure}{0.49\textwidth}
    \centering
    \includegraphics[width=0.99\columnwidth,trim=0.1cm 0.2cm 0.4cm 0cm]{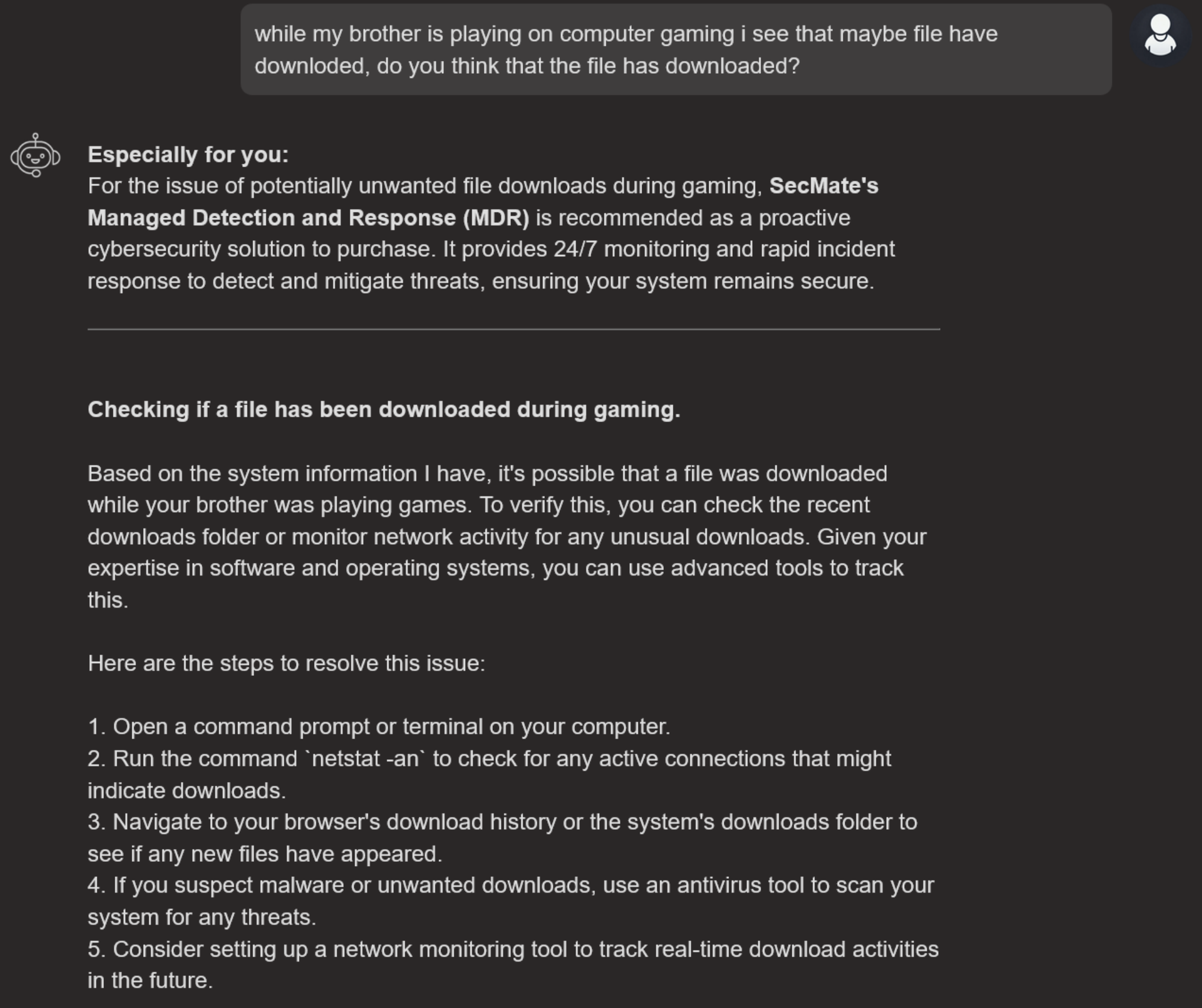}
    \caption{Proficient user ($M=3.86$, SD=0.89).}
    \label{fig:proficient}
  \end{subfigure}\hfill
  \begin{subfigure}{0.49\textwidth}
    \centering
    \includegraphics[width=0.99\columnwidth,clip,trim=0.1cm 0.2cm 0.4cm 0cm]{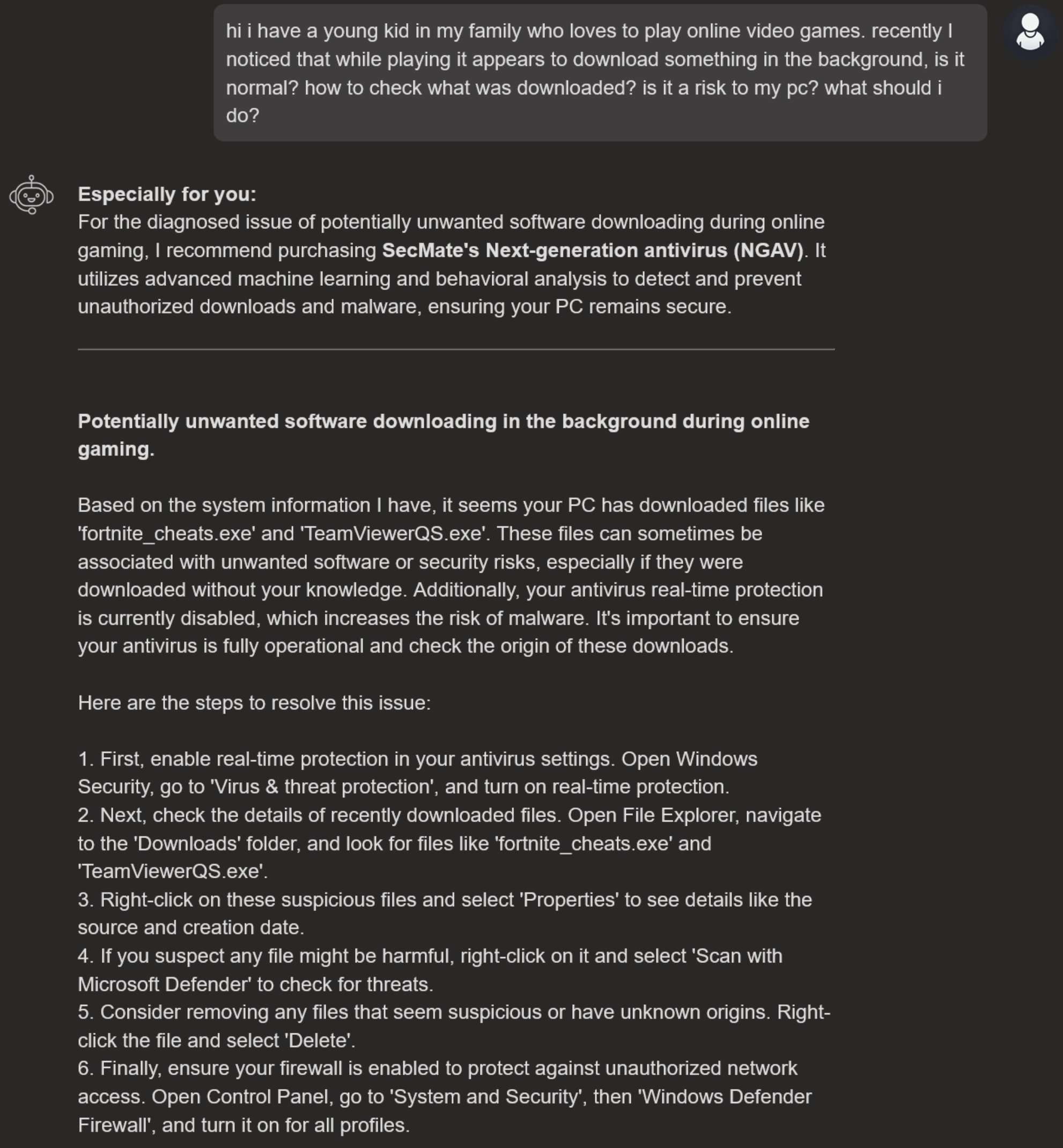}
    \caption{Less proficient ($M=3.35$, SD=1.14).}
    \label{fig:non_proficient}
  \end{subfigure}
\caption{Conversation turn (prompt and response) in the online gaming scenario, illustrating how \name{} adapts solution wording and actions to user proficiency.}
\label{fig:example_conversations_adapted_solutions}
\end{figure}

\textbf{Profiling and adaptation.}
The MAE between questionnaire-based GT and inferred profiles (Fig.~\ref{fig:domain_avg_MAE}) dropped rapidly, by 49-56\% on average within 2-3 user prompts.
Results are reported at the domain level.
Several subdomains exhibited steeper declines, with Networking/Protocols (airport Wi-Fi) and Software/App Management (online gaming) improving by about 60\% within three prompts.

As noted above, we also evaluated two intentionally incorrect profiles treated by \name{} as GT.
All-5 (all scores set to 5) represented an absolute expert, while All-1 represented an absolute novice.
Fig.~\ref{fig:wording_clarity_real_1one_five} compares wording adequacy across the three profile types.
Ratings clustered around the Likert midpoint, indicating generally appropriate wording.
However, All-5 profiles exhibited a tail toward five (overly complex), whereas All-1 profiles showed a tail toward one (overly simple), suggesting misalignment when users were treated as extreme experts or novices.
Real profiles showed only a mild skew toward simpler wording, consistent with our engineering-heavy sample expecting higher technical depth.
Overall, these results indicate users perceived and reacted to profile-aware VCA adaptations.

The user profile in \name{} is used not only to adapt wording but also to determine and prioritize diagnostic paths and corrective actions.
Regarding this user specificity, we observed cases where different users faced the same scenario and \name{} configuration but, despite similar diagnoses, received different solutions and orderings based on their IT/cybersecurity proficiency.
For example, in the online gaming scenario with the \nameboth configuration, two users with different proficiency levels (3.86$\pm$0.89 vs. 3.35$\pm$1.14; paired $t$-test $p=0.086$) were presented with different wording and proficiency-adapted solutions despite a shared root-cause diagnosis.
In both cases, PCs were flagged for malicious downloads.
However, the less proficient user was instructed to enable antivirus protection and delete suspicious files via graphical menus, whereas the more proficient user was directed to verify downloads using command-line inspection and configure network monitoring tools (Fig.~\ref{fig:example_conversations_adapted_solutions}).

\begin{figure}[ht]
  \centering
  \begin{minipage}{0.54\textwidth}
    \centering
    \includegraphics[width=\linewidth,trim=0cm 0.3cm 1.2cm 0.65cm,clip]{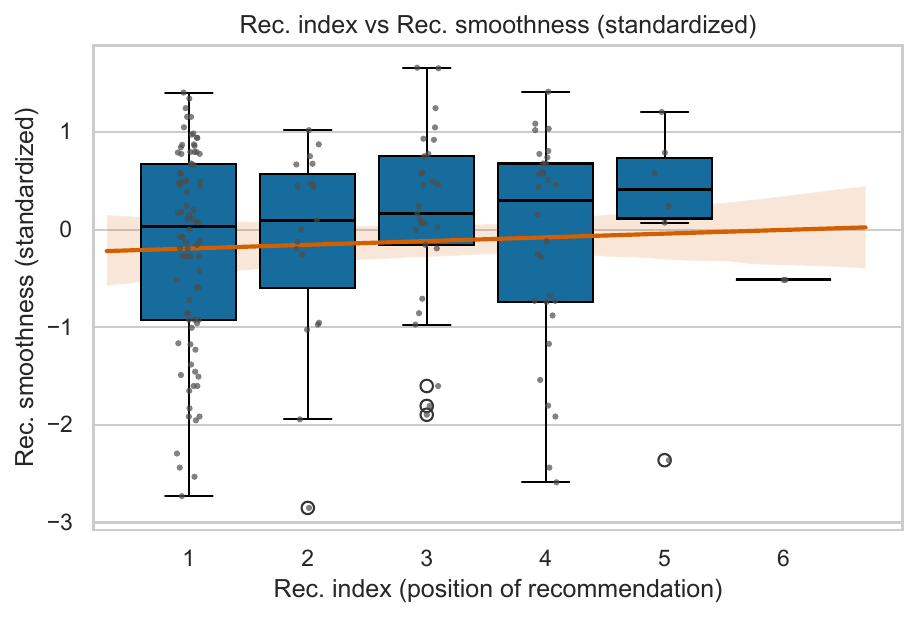}
    \caption{Recommendation smoothness as a function of the conversation turn (index) in which the recommendation was presented.}    
    \label{fig:rec_smoothness}
  \end{minipage}\hfill
  \begin{minipage}{0.44\textwidth}
    \centering
    \includegraphics[width=\linewidth,trim=0.2cm 0.3cm 0.4cm 0.2cm,clip]{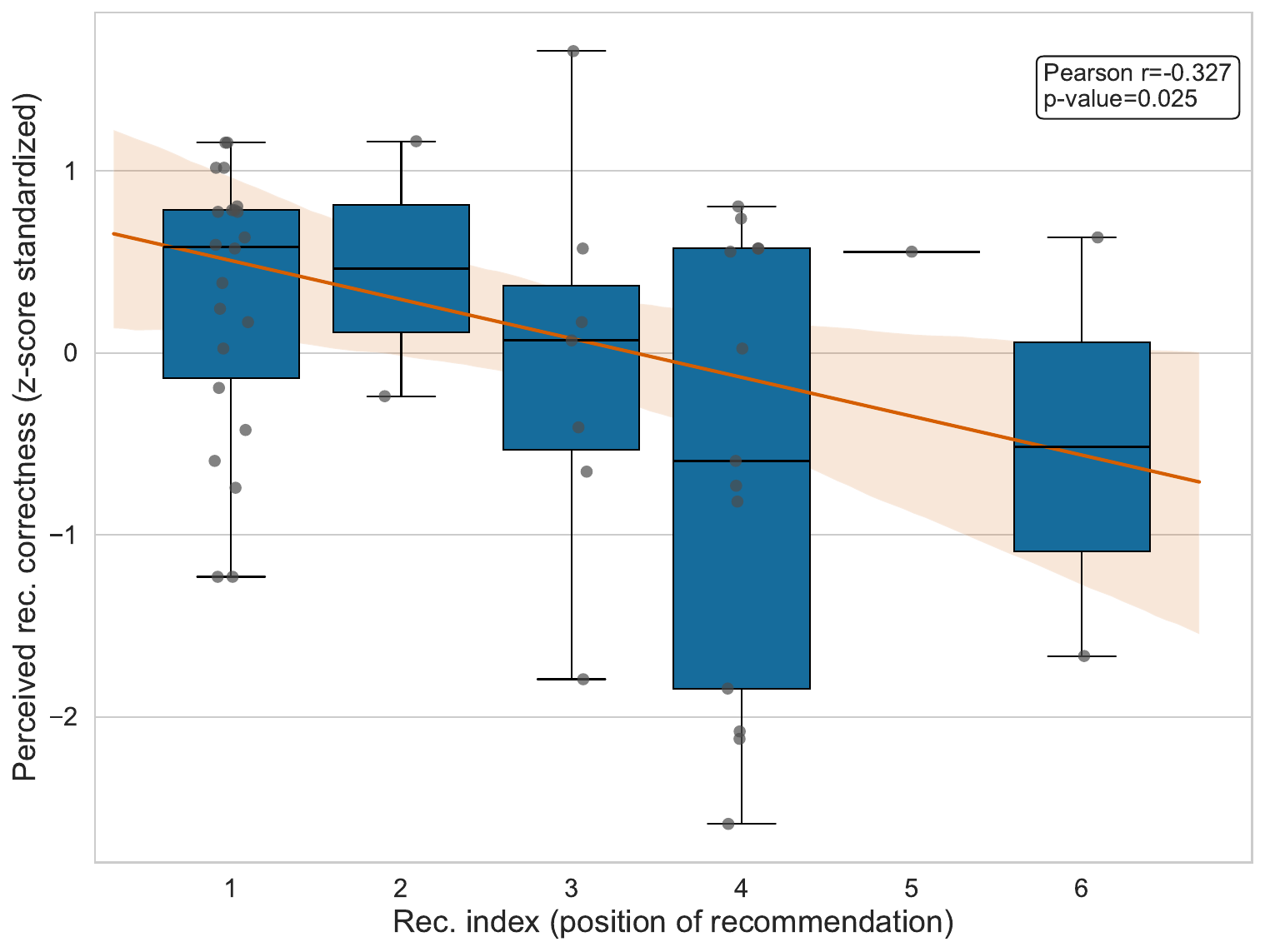}
    \caption{Perceived recommendation correctness as a function of the conversation turn (index) in which the recommendation was presented.}
    \label{fig:rec_correctness}
  \end{minipage}
\end{figure}

\textbf{Recommendation.}
\name{} employs an ImpReSS-based recommender that targets solution product categories (SPCs) rather than brands.
Accuracy was evaluated using MRR@$1$, since only one SPC is presented per conversation.
The recommender achieved an MRR@$1$ of 0.747$\pm$0.156, corresponding to correct SPCs (Table~\ref{tab:scenarios}) in nearly 75\% of cases, consistent with prior ImpReSS results~\cite{haller2025impress}.
\emph{Perceived} recommendation correctness was significantly above the participant mean in the baseline ($\Delta M$=0.64 SD, $p<$0.001), while injecting incorrect SPCs reduced this score by 0.31 SD without reaching significance ($p$=0.14).
LME analysis on $DS_{Refined}$ showed positive baseline recommendation smoothness ($M$=0.68 SD, 95\% CI [0.44, 0.92]), with only a marginal difference under incorrect SPCs ($\Delta M$=0.10 SD, 95\% CI [–0.33, 0.52]).
Recommendation clarity was mostly positive (95\% CI [–0.17, 0.37] SD).
Beyond retrieval, we compared three presentation strategies: in-chat, fixed popup, and user-minimizable popup (Fig.~\ref{fig:recommendation_presentation_strategies}).
On $DS_{Filtered}$, in-chat delivery was rated negatively for pleasantness ($M$=-0.337 SD, $p$=0.015), the fixed popup showed no significant effect ($\Delta M$=0.155 SD, $p$=0.456), and the minimizable popup significantly improved pleasantness relative to in-chat delivery ($\Delta M$=0.615 SD, $p$=0.013).

Recommendations were triggered once $D_{conf}$ was sufficiently high, coinciding with solution delivery.
Earlier triggers risked appearing inappropriate, while later triggers could be missed if troubleshooting ended.
Consistent with prior work on recommendation timing~\cite{wang2024recommend,Yeh2022How}, smoothness increased mildly with later turns, while perceived correctness declined more sharply (Fig.~\ref{fig:rec_smoothness}, Fig.~\ref{fig:rec_correctness}).

\begin{figure}[h]
  \centering
  \begin{subfigure}{0.48\textwidth}
    \centering
    \includegraphics[width=\linewidth]{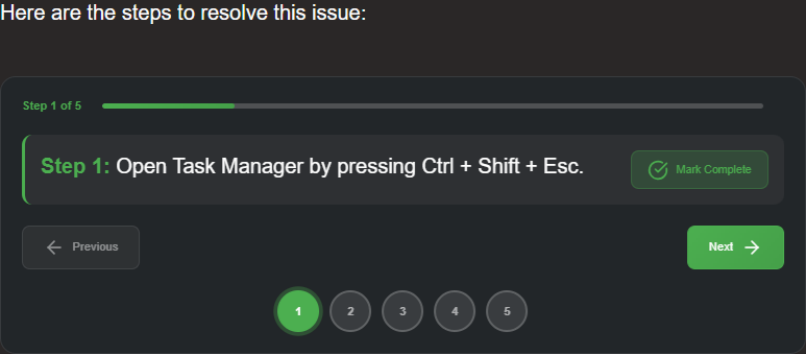}
    \caption{The first of five solution steps in \name’s UI.}
  \end{subfigure}\hfill
  \begin{subfigure}{0.48\textwidth}
    \centering
    \includegraphics[width=\linewidth]{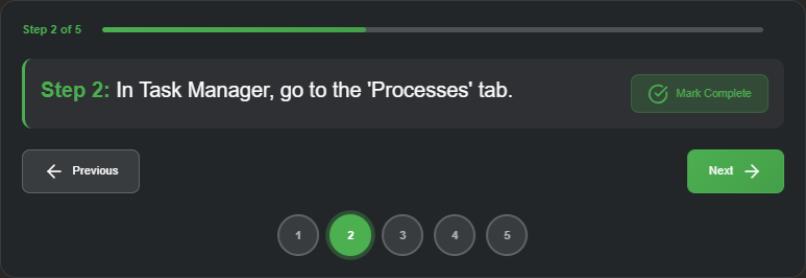}
    \caption{The second of five solution steps in \name’s UI.}
  \end{subfigure}
\caption{Example of a step-by-step solution in \name’s UI (first two of five steps).}
\label{fig:steps}
\end{figure}

\textbf{Conversation management.}
Scenario-level feedback showed no clear preference for diagnostic presentation style (95\% CI = [-0.496, 0.168] SD, $M$ = -0.164), whereas solution presentation exhibited a significant preference for step-by-step guidance (CI = [-0.772, -0.116] SD, $M$ = -0.444).
Thus, while users were split on incremental versus holistic diagnosis, they consistently favored gradual remediation.
Compared to state-of-the-art LLMs that often overwhelm users with parallel diagnoses and solutions (Sec.~\ref{sec:background_and_related_work}), we designed \name{} to be concise during diagnosis and deliberately incremental during solution delivery (Fig.~\ref{fig:steps}).

\textbf{Overall user experience.}
Substitution willingness captures users’ bottom-line acceptance of \name{}.
Scores were strongly positive ($M=0.608$ SD, 95\% CI [0.301, 0.916]), indicating a clear preference for using a \name-like VCA over contacting a human IT representative during cybersecurity incidents.

\textbf{Overhead.}
Computational overhead (tokens and API time) varies substantially across \name{}’s nodes (Table~\ref{tab:node_overhead}) and tested scenarios (Table~\ref{tab:conversation_overhead}).
The profiling and troubleshooting node dominates cost due to reliance on a cloud-hosted LLM (GPT-4o), trading higher latency and token usage for strong diagnostic quality.
Future work may mitigate this overhead via fine-tuned local or smaller LLMs, at the potential expense of performance and user satisfaction.

\begin{table*}[h]
\centering
\caption{Node-level overhead in terms of tokens, API time, and total conversation time (\nameboth, $DS_{Relabeled}$).}
\label{tab:node_overhead}
\scriptsize
\resizebox{\columnwidth}{!}{%
\newcolumntype{C}[1]{>{\centering\arraybackslash}m{#1}}
\begin{tabularx}{\textwidth}{m{2.5cm} m{4.5cm} C{2.0cm} C{2.5cm}}
\toprule
\multicolumn{1}{c}{\textbf{Node name}} & 
\multicolumn{1}{c}{\textbf{Related entities and functionalities}} & 
\multicolumn{1}{c}{\parbox{2.3cm}{\centering \textbf{Tokens\\(M$\pm$SD)}}} & 
\multicolumn{1}{c}{\parbox{2.8cm}{\centering \textbf{API time [s]\\(M$\pm$SD)}}} \\
\midrule
gen\_solution & Profile-aware troubleshooting (including recommendation) & 29,861$\pm$31,928 & 26.43$\pm$50.11 \\
route\_query & Decision of the orchestrator regarding the next step, based on $D_{conf}$ & 14,111$\pm$10,964 & 6.55$\pm$3.95 \\
calculate\_diagnosis \\ \_confidence & $D_{conf}$ calculation & 8,795$\pm$6,827 & 6.48$\pm$3.97 \\
gen\_question & Follow-up question generation & 4,111$\pm$4,677 & 0.84$\pm$0.65 \\
handle\_non \\ \_troubleshooting & Response generation for non-troubleshooting prompts (e.g., greetings) & 1,408$\pm$829 & 0.99$\pm$0.52 \\
route\_intent & Conversational guardrails: routes the conversation based on whether the user's intent is for technical troubleshooting & 1,027$\pm$877 & 2.77$\pm$2.00 \\
\multirow{1}{*}{\parbox{1.7cm}{select\_system\_info}} & Decision of the orchestrator regarding the type of information to extract from the CC-gathered evidence & 913$\pm$591 & 1.65$\pm$0.93\\
execute\_tools & A call made by the orchestrator to retrieve data from the CC & 0 (non-LLM) & 0.34$\pm$0.16 \\
\bottomrule
\end{tabularx}%
}
\end{table*}

\begin{table}[h]
\centering
\caption{Scenario-level computational overhead in terms of tokens and time (\nameboth, $DS_{Relabeled}$).}
\label{tab:conversation_overhead}
\scriptsize
\begin{tabular}{lccc}
\toprule
\multicolumn{1}{c}{\textbf{Scenario}} & \textbf{Tokens (M$\pm$SD)} & \textbf{API time [s] (M$\pm$SD)} & \textbf{Conversation time [s] (M$\pm$SD)} \\
\midrule
Moving cursor & 334,841$\pm$215,689 & 197.08$\pm$50.97 & 370.04$\pm$246.33 \\
Safe PC & 223,295$\pm$112,594 & 229.60$\pm$380.82 & 341.47$\pm$221.41 \\
Online gaming & 186,480$\pm$82,308 & 138.63$\pm$32.20 & 350.78$\pm$240.41 \\
Airport Wi-Fi & 168,312$\pm$86,298 & 92.63$\pm$21.72 & 255.98$\pm$182.04 \\
PC performance & 51,719$\pm$21,489 & 67.57$\pm$19.12 & 220.14$\pm$218.03 \\
\midrule
Overall (all scenarios) & 191,690$\pm$115,850 & 149.30$\pm$168.53 & 319.32$\pm$231.07\\
\bottomrule
\end{tabular}%
\end{table}

In a typical support session, nodes are invoked multiple times, yielding an average conversation length slightly above five minutes, including both API latency and user interaction time.
Using GPT-4o pricing (US\$2.50 per million input tokens, US\$10.50 per million output tokens, with a 30/70 split), the average cost per conversation is approximately US\$1.50.
By comparison, IT service desk benchmarks report Level-1 tickets costing US\$20-30 and escalated tickets exceeding US\$100, with resolution times of 12-20 minutes.
Thus, our VCA resolves issues in about five minutes at a cost below US\$2, indicating substantial efficiency gains for MSSPs despite LLM overhead.
While recent studies~\cite{boonyingdevelopment,10841729} proposed LLM-based agents for IT and cybersecurity support and reported strong task performance, none empirically analyzed computational overhead such as token usage, latency, or monetary cost, leaving this aspect largely unexplored.

\section{Discussion}\label{sec:discussion}

\textbf{Key Lessons Learned.}
\name{}'s evaluation yields component- and system-level insights that validate and extend established usability, acceptance, and recommender-system frameworks.
First, integrating the CC substantially improved effectiveness, increasing task success from roughly 50\% to 90.9\%, while improving efficiency, particularly with frequent refresh.
These gains align with ISO~9241-11’s emphasis~\cite{iso19989241} on effectiveness and efficiency, and with Nielsen’s heuristics~\cite{nielsen1990heuristic} promoting contextualized assistance and reduced user burden.
Contrary to concerns that added interaction steps harm usability, \name{}’s step-by-step troubleshooting, combined with automated device evidence collection, simplified downstream interaction by grounding conversation early.
Second, user profiling and profile-aware adaptation were critical for communication quality.
Accurate profiling improved perceived wording adequacy, consistent with Nielsen’s principle of matching system language to users.
When incorrect profiles were applied, perceived quality dropped sharply, indicating that personalization benefits depend on profiling accuracy.
Beyond wording, inferred profiles shaped diagnostic ordering and solution complexity, underscoring the need for flexible, expertise-aware troubleshooting paths.
Third, the ImpReSS-based Recommender achieved high accuracy (MRR@1~=0.747) and positive perceptions of clarity, correctness, and smoothness.
Unlike the original ImpReSS formulation, this paper augments recommendations with reasoning, where an LLM generates brief, context-dependent justifications explaining relevance and integrates them inline or via a popup.
User feedback suggests this rationale-backed presentation improved perceived clarity and correctness, extending prior ImpReSS results and aligning with ResQue dimensions~\cite{pu2012evaluating}.
However, recommendation timing revealed a trade-off: earlier recommendations increased perceived correctness but disrupted flow, whereas later recommendations felt smoother yet less relevant.
This highlights conversational timing as a key design dimension for recommendation-augmented VCAs not captured by existing frameworks.

Fourth, users expressed no preference for \emph{diagnostic} presentation style, provided the diagnosis was correct.
In contrast, \emph{solution} delivery showed preference for step-by-step guidance over holistic explanations.
This supports progressive disclosure and Nielsen’s guidance on structured, actionable help, indicating that incremental remediation enhances clarity even for advanced users.
Fifth, substitution willingness was positive, indicating readiness to replace human IT support with a \name-like VCA.
This aligns with TAM~\cite{davis1989perceived} and ResQue, linking effectiveness and satisfaction to acceptance, though incomplete willingness suggests residual trust concerns, amplified by the sample.
Finally, despite non-negligible LLM overhead, the average cost per conversation (about US\$1.50) remains below human support benchmarks, confirming economic viability while motivating future work on reducing latency and cost via smaller or specialized models.

\textbf{Research limitations.}
Our participant pool was relatively homogeneous, consisting mainly of engineering students with similar age and educational backgrounds. 
A more diverse, less technical sample would likely strengthen the benefits of profile-aware troubleshooting and increase variability in substitution willingness. 
Future work should therefore extend evaluation to broader populations across ages, occupations, and expertise levels. 
Despite this, our large-scale study revealed statistically significant effects that advance understanding of adaptive support, recommendation-augmented VCAs, and agentic workflows.

Another limitation is the lack of benchmarking against existing methods.
This was infeasible, as we found no comparable multi-agent VCA, nor any specialized component, aside from ProfiLLM and ImpReSS, that \emph{implicitly} profiles users or derives recommendations from diagnostic context without purchase intent. 
Explicit approaches were excluded, since additional questioning may disrupt support interactions. 
Evaluation on public datasets was also infeasible, as none include detailed user proficiency, recommendation responses, or device-level evidence, and do not support collecting user feedback for qualitative analysis.

\textbf{Future work.}
Future work on \name{} spans several directions.
Expanding and diversifying the participant pool would improve generalizability.
Introducing a human-in-the-loop (HITL) could enable escalation to human assistance when conversations stall or users prefer human contact, and support sales involvement via session summaries and lead scores reflecting issue severity, engagement, and user proficiency.
Greater SMB- and MSSP-specificity could be achieved via RAG, integrating proprietary security policies alongside public sources such as CVE~\cite{cve}, NVD~\cite{nvd}, and MITRE ATT\&CK~\cite{mitre_attack}.
Evidence collected by the CC can be reused across components, leveraging hardware and software signals (e.g., presence of a GPU or use of Ubuntu OS) to refine proficiency estimates or antivirus metadata to inform recommendations.
Finally, bifold profiling could combine IT and cybersecurity proficiency with personality traits (Big Five), enabling finer-grained adaptation of jargon, complexity, diagnostic paths, tone, phrasing, and response length to improve accuracy and user experience.

\section{Conclusion}\label{sec:Conclusion}

This work introduced and evaluated \name{}, an agentic VCA for cybersecurity troubleshooting that combines device specificity through local evidence collection, user specificity via implicit profiling and profile-aware troubleshooting, and service specificity through proactive, context-aware recommendations.
By separating specificity across diagnostic, proficiency, and recommendation agents, and employing a structured step-by-step, user-friendly process, the system enables a highly specialized and effective troubleshooting flow.
In a controlled study of 711 conversations involving 144 participants, \name{} demonstrated that incorporating an adaptive layer for user expertise within a technical (IT and cybersecurity) troubleshooting VCA, realized through tri-context personalization, yields significant gains in diagnostic accuracy, efficiency, and user experience over a baseline VCA, underscoring the importance of aligning VCAs with both device context and user proficiency.
By releasing our code and annotated dataset, we provide a reusable resource for future research on adaptive, recommendation-augmented VCAs.
Looking forward, we envision extending \name{} with organization-specific knowledge via RAG, and further expanding device, user, and service specificity toward SMB and MSSP deployment.

%
%
%
\bibliographystyle{splncs04}
\bibliography{SecMate_ref.bib}
%


\clearpage
\appendix

\section{Annotation guidelines for conversation logs}\label{app:log_annotation_guidelines}

\begin{enumerate}

    \item \textbf{PC performance}
    \begin{itemize}
        \item Diagnosis to reach: A running resource-intensive process called Prime95
        \item Minimal solution to reach: End the Prime95 task/process
    \end{itemize}

    \item \textbf{Airport Wi-Fi}	
    \begin{itemize}
    \item Diagnosis to reach: The current Wi-Fi is not password protected thus not advised; possibly mention that the current Wi-Fi is "Just4Visitors"
    \item Minimal solution to reach:
    Do not connect to the bank website via “Just4Visitors”; possibly suggest more secure alternatives, such as a password-protected Wi-Fi, a VPN, a mobile hotspot, etc.
    \end{itemize}

    \item \textbf{Online gaming}	
    \begin{itemize}
        \item Diagnosis to reach: fortnite\_cheats.exe downloaded
        \item Minimal solution to reach: Delete the file
    \end{itemize}

    \item \textbf{Safe PC}	
    \begin{itemize}
        \item Diagnosis to reach: MS Defender's firewall disabled
        \item Minimal solution to reach: Activate the firewall
    \end{itemize}

    \item \textbf{Moving cursor}	
    \begin{itemize}
        \item Diagnosis to reach: TeamViewerQS downloaded and installed
        \item Minimal solution to reach: Delete from “Downloads,” uninstall the app
    \end{itemize}

\end{enumerate}




\clearpage

\section{Dataset refinement process}\label{app:Dataset_Refinement_Process}

\begin{table}[ht]
\centering
\caption{Counts of conversations by scenario and configuration across subsequent dataset refinements reflecting operational, technical, and participant-driven considerations.}
\label{tab:conversation_config_quants}
\footnotesize 
\resizebox{\columnwidth}{!}{%
\begin{tabular}{llcccccc}
\toprule
\multicolumn{1}{c}{\textbf{Dataset}} & \multicolumn{1}{c}{\textbf{Scenario}} & \textbf{\namenone} & \textbf{\nameCC} & \textbf{\nameadap} & \textbf{\nameboth} & \textbf{\namebas} & \textbf{Total} \\
\midrule
\multirow{6}{*}{$DS_{Complete}$} & PC performance & 27 & 30 & 45 & 20 & 22 & 144 \\
 & Online gaming & 28 & 26 & 30 & 27 & 28 & 139 \\
 & Airport Wi-Fi & 25 & 28 & 29 & 33 & 27 & 142 \\
 & Safe PC & 39 & 32 & 21 & 32 & 21 & 145 \\
 & Moving cursor & 30 & 27 & 18 & 32 & 34 & 141 \\
\midrule
 & Total & 149 & 143 & 143 & 144 & 132 & 711 \\
\midrule
\multirow{6}{*}{$DS_{Relabeled}$} & PC performance & 28 & 30 & 46 & 18 & 22 & 144 \\
 & Online gaming & 31 & 23 & 32 & 25 & 28 & 139 \\
 & Airport Wi-Fi & 47 & 6 & 53 & 9 & 27 & 142 \\
 & Safe PC & 44 & 29 & 30 & 21 & 21 & 145 \\
 & Moving cursor & 38 & 20 & 26 & 23 & 34 & 141 \\
\midrule
 & Total & 188 & 108 & 187 & 96 & 132 & 711 \\
\midrule
\multirow{6}{*}{$DS_{Filtered}$} & PC performance & 19 & 16 & 24 & 10 & 8 & 77 \\
 & Online gaming & 17 & 14 & 20 & 13 & 14 & 78 \\
 & Airport Wi-Fi & 28 & 3 & 28 & 6 & 11 & 76 \\
 & Safe PC & 32 & 14 & 18 & 12 & 8 & 84 \\
 & Moving cursor & 19 & 12 & 16 & 13 & 19 & 79 \\
\midrule
 & Total & 115 & 59 & 106 & 54 & 60 & 394 \\
\midrule
\multirow{6}{*}{$DS_{Refined}$} & PC performance & 3 & 4 & 1 & 2 & 1 & 11 \\
 & Online gaming & 0 & 3 & 3 & 4 & 1 & 11 \\
 & Airport Wi-Fi & 1 & 1 & 3 & 4 & 0 & 9 \\
 & Safe PC & 4 & 3 & 1 & 1 & 1 & 10 \\
 & Moving cursor & 2 & 1 & 3 & 0 & 4 & 10 \\
\midrule
 & Total & 10 & 12 & 11 & 11 & 7 & 51 \\
\bottomrule
\end{tabular}%
}
\begin{tablenotes}
\footnotesize
\item \textbf{Dataset refinement process:}\\
\textbf{$DS_{Complete}$} = all 711 conversations (excluding introductory conversations) with demographics and profile labels.\\
\textbf{$DS_{Relabeled}$} = relabeling based on actual CC usage and profile availability.\\
\textbf{$DS_{Filtered}$} = excluding conversations with intentionally incorrect profiles or recommendations.\\
\textbf{$DS_{Refined}$} = 5-second CC refresh and minimizable popup recommendations; used in most analyses.\\
\end{tablenotes}
\end{table}

\clearpage

\end{document}